\documentclass[10pt,conference]{IEEEtran}

\usepackage{verbatim}  
\usepackage{graphicx}
\usepackage{url}
\usepackage{amssymb}
\usepackage{amsmath}
\usepackage{subfigure}
\usepackage{times}
\usepackage{cite}

\newcommand{\urlsamefont}[1]
{
\urlstyle{same}\url{#1}
}

\def\maxoc{maximum number of outgoing connections}

\def\baselinestretch{0.99}

\begin{document}
\title{Swarming Overlay Construction Strategies}
\author{
\IEEEauthorblockN{Anwar Al-Hamra}
\IEEEauthorblockA{Hariri Canadian University\\
Damour, Lebanon\\
hamraaw@hcu.edu.lb}
\and
\IEEEauthorblockN{Nikitas Liogkas\IEEEauthorrefmark{1}}
\IEEEauthorblockA{Knight Equity Markets, L.P.\\
Jersey City, NJ, USA\\
nliogkas@knight.com}
\IEEEauthorblockA{\IEEEauthorrefmark{1}This work was completed while the\\
author was a Ph.D. student at UCLA.}
\and
\IEEEauthorblockN{Arnaud Legout, Chadi Barakat}
\IEEEauthorblockA{I.N.R.I.A.\\
Sophia Antipolis, France\\
\{arnaud.legout, chadi.barakat\}@sophia.inria.fr}
}

\maketitle

\begin{abstract}
  Swarming peer-to-peer systems play an increasingly instrumental role
  in Internet content distribution. It is therefore important to
  better understand how these systems behave in practice. 
  Recent research efforts have looked at various 
  protocol parameters and have measured how they affect
  system performance and robustness. However, the importance of the strategy 
  based on which peers establish connections
  has been largely overlooked.

  This work utilizes extensive simulations to examine 
  the default overlay construction strategy in BitTorrent systems.
  Based on the results, we identify a critical parameter, the maximum allowable
  number of outgoing connections at each peer, and evaluate its impact
  on the robustness of the generated overlay.
  We find that there is no single optimal value for this parameter using the default 
  strategy.
  We then propose an alternative strategy that allows certain new peer connection 
  requests to replace existing connections.
  Further experiments with the new strategy demonstrate that it outperforms the default 
  one for all considered metrics by creating an overlay more robust to churn.
  Additionally, our proposed strategy exhibits optimal behavior for a well-defined
  value of the maximum number of outgoing connections, thereby removing the need to set 
  this parameter in an ad-hoc manner\footnote{In proc. of ICCCN'2009,
    San Francisco, CA, USA. August 2009. (author version).\\ \copyright2009 IEEE. Personal use of this material is
  permitted. However, permission to reprint/republish this material
  for advertising or promotional purposes or for creating new
  collective works for resale or redistribution to servers or lists,
  or to reuse any copyrighted component of this work in other works
  must be obtained from the IEEE.}.
\end{abstract}

\begin{IEEEkeywords}
BitTorrent, overlay construction, preemption, robustness, outgoing connections
\end{IEEEkeywords}

\section{Introduction}

Recent research efforts have examined the protocol parameters of popular swarming 
peer-to-peer content distribution systems, in order to identify their impact
on system performance and robustness. Such efforts have mainly focused 
on the protocol algorithms that are believed to be the major factors
affecting system behavior, such as BitTorrent's piece and peer selection
strategies~\cite{bharambe06,felber04,lego06_IMC,lego07_SIGMETRICS,TIAN06_INFOCOM}. 

However, the actual manner by which peers form connections and the overlay is 
constructed has been largely overlooked. 
As shown by Urvoy \textit{et al.}~\cite{URVO05}, the time needed to distribute 
content in BitTorrent is directly affected by the overlay topology. 
Moreover, Ganesh \textit{et al.}~\cite{ganeshInfocom2005} evaluated the impact 
of the overlay structure on the spread of epidemics, which can be viewed as a 
special case of robustness in peer-to-peer file replication. To the best 
of our knowledge, there has been no study that specifically investigated optimal
overlay construction strategies for content replication.

In this paper, we evaluate two such strategies in the BitTorrent
protocol.  We first present and evaluate the \textit{tracker
  strategy}, which most BitTorrent implementations use by default to
guide new connection establishment.  We identify a concrete
shortcoming, namely the strategy's tendency to cause peer clustering
and potential network partitions, which might have an adverse impact
on system robustness.  To address this, we introduce an alternative,
the \textit{preemption strategy}, which dictates giving preference to
certain new peer connection requests.  We evaluate the properties of
overlays generated by both strategies using extensive simulations,
focusing on flash crowd scenarios, when the system is under high load
and more vulnerable to churn.

Indeed, the flash crowd phase as it is the most critical phase for a
torrent, as there is a single seed. In case some peers become
disconnected from this initial seed, they will experience a much
higher download completion time. Moreover, a poorly structured overlay
may result in a slower propagation of the pieces, thus a lower overall
performance. However, in this study, we focus on the overlay
property rather than on its impact on performance.

Based on our results, we identify the \textit{maximum number of outgoing connections} 
as a parameter that significantly affects the structure and properties of the generated 
overlay. This parameter is currently used in BitTorrent to enforce a hard upper limit 
on the number of connections a peer can initiate.
We define metrics that characterize the overlay structure, and compute
these metrics for various values of the number of maximum outgoing connections per peer.

The contributions of this work include the following.
\begin{enumerate}
\item We show that, for the default BitTorrent overlay construction strategy, there is 
no single value of the maximum number of outgoing connections that optimizes all considered  
metrics. In addition, a value between $20$ and $30$ clearly offers a better 
choice than the usual default value of $40$.
\item We also show that our proposed preemption strategy outperforms the default one for 
all metrics. For this strategy, a maximum number of outgoing connections that is 
simply equal to the maximum peer set size (number of neighbors) presents the best choice. As a result, our proposed strategy, while simple and easy to implement, 
removes the need to set the maximum number of outgoing connections in an ad-hoc 
manner, thereby simplifying the protocol.
\end{enumerate}
 
The rest of this paper is organized as follows. In
Section~\ref{sec:terminology} we define the terms we use.
We present the tracker and preemption strategies in Section~\ref{sec:strategies}.
Section~\ref{sec:methodology} then describes our experimental
methodology, while our results on the properties of overlays generated
with both strategies are presented in Section~\ref{sec:results}. 
Section~\ref{sec:related} describes related work, and 
we conclude and outline future work in Section~\ref{sec:conc}.

\section{Terminology}
\label{sec:terminology}
In this section, we present the terms used to describe the BitTorrent Overlays. 

\noindent
\textbf{Peer Set:} Each peer maintains a list of other peers to
  which it has open TCP connections. This list is called the peer set, also known
  as the neighbor set. Thus, a \textit{neighbor} of peer $P$ is a peer that 
  belongs to $P$'s peer set.

\noindent
\textbf{Maximum Peer Set Size:} The upper limit on the number of peers that 
  can be in the peer set. It is a configuration parameter of the protocol.

\noindent
\textbf{Average Peer Set Size:} A torrent-wide metric calculated by summing up
  the peer set size for each peer in the torrent, and dividing by the total number
  of peers.

\noindent
\textbf{Incoming and Outgoing Connections:} When a peer $A$ initiates a TCP connection
  to peer $B$, we say that $A$ has an \textit{outgoing} connection to $B$, and that $B$ 
  has accepted an \textit{incoming} connection from $A$. Note that all connections are 
  really bidirectional, they are just flagged as incoming or outgoing. This flag has no 
  impact on the actual data transfer, however, it is used to decide whether a new outgoing
  connection can be established, as explained in Section~\ref{sec:bittorrent-overview}.

\noindent
\textbf{Maximum Number of Outgoing Connections:} The upper limit on the number of
  outgoing connections a peer can establish. This is a configuration parameter of the 
  protocol. 

\section{Overlay Construction Strategies}
\label{sec:strategies}
We first present the overlay construction strategy BitTorrent follows, 
then propose an alternative based on preempting existing connections.

\subsection{Tracker Strategy}
\label{sec:bittorrent-overview}

A piece of content to be distributed with BitTorrent is first divided into 
multiple pieces. A \textit{metainfo file} is then created by the content provider,
which contains all the information necessary for the download, including the 
number of pieces, \mbox{SHA-1} hashes that are used to verify the integrity of 
received data, and the IP address and port number of the tracker.
To join a torrent, a peer retrieves the metainfo file 
out of band, usually from a well-known Web site. $P$ then contacts
the tracker who returns a random subset of other peers already participating
in the download; we call this subset the \textit{initial peer set}. 
A typical number returned by many tracker implementations is 
$80$, which is also what we use for our simulations. 

After receiving this initial
peer set, the new peer attempts to initiate new connections,
under the following two constraints: 1) a peer is not allowed to establish
more than a fixed number of outgoing connections, typically $40$, and
2) a peer cannot maintain in total more than a fixed number of open
connections, typically $80$ (the maximum peer set size). 
The latter limit is imposed to avoid performance degradation due to competition
among TCP flows, while the former serves to ensure that some connection slots
are kept open for new peers that will join later. In this manner, 
the initial peer set can be augmented later by connections initiated by remote
peers.

Whenever the peer set size falls below a given threshold (typically $20$),
a peer contacts the tracker again and asks for more. To avoid overwhelming
the tracker with such requests, there is usually a minimum interval between two 
such consecutive messages.
Finally, each peer contacts the tracker periodically (typically once every 
$30$ minutes) to indicate that it is still present in the network. If no heartbeat
is received for more than $45$ minutes, the tracker assumes the peer has 
left the system, and does not include it in future initial peer sets.

\subsection{Preemption Strategy}
\label{sec:track-with-preempt}

The potential shortcoming of the default strategy can be seen by considering
the effect of the maximum number of outgoing connections. A small number for this parameter
will allow peers who have recently joined the system to connect to older ones, whereas a 
large number will cause peers to be more connected to others that joined around the same time. 
Thus, we expect that, when increasing this value, we will observe the formation of clusters
of peers that joined close together in time. For very large values, close to the maximum peer
set size, this could even cause the creation of mostly disjoint cliques that share data within
themselves, thereby compromising the robustness of the system to churn. If the connecting
peer between two cliques were to disconnect, we would have the creation of partitions in the
system. Our results bear out this hypothesis.

To address this issue, we propose an alternative strategy based on preempting existing 
connections.
The only difference from the default strategy manifests itself when a peer $A$ wants to 
establish a connection to a peer $B$ that has already reached its maximum peer set size. 
In the default strategy such a connection attempt would simply be rejected. With preemption, 
however, peer $B$ will accept the new connection after dropping an existing one, if and
only if $A$ has discovered $B$ from the tracker (as opposed to through other means,
e.g., peer exchange).

Thus, an implementation of the preemption strategy would be exactly the same as the
default one, with the following modification. When peer $P_j$ joins a torrent, it receives 
the IP addresses of several existing peers including peer $P_i$. Let us assume
that $P_j$ attempts to initiate a connection to peer $P_i$. If $P_i$ has not
reached its maximum peer set size, the connection is accepted with no further 
action. However, if $P_i$ has already reached its maximum peer set
size, it will either 1) accept the connection from $P_j$, after tearing down an
existing connection, if $P_j$ discovered $P_i$'s IP address
from the tracker, or 2) refuse the connection in any other case.

The rationale behind this strategy is the goal of introducing some randomness in the 
connection establishment process, to help convergence to the maximum peer set
size as fast as possible and prevent cliques. The default strategy gives preference to 
connections from peers who joined close in time, especially at the beginning of the download. 
The preemption strategy attempts to spread connections uniformly over the peers delivered 
in the peer lists by the tracker, without being affected by external peer connection 
mechanisms (e.g., peer exchange).

Note that, if $P_i$ decides to accept the new connection, it selects the connection to 
close at random among all the connections that were initiated by remote peers
(the incoming connections). In case there is no such connection, it selects any 
connection at random.
The rationale behind this is to maximize the probability that the remote peer can quickly 
recover from such an unexpected connection drop. 
Indeed, in case the remote peer $P_i$ has reached its
maximum number of outgoing connections, closing an incoming
connection from peer $P_k$ (an outgoing connection for $P_k$) will allow $P_k$ to 
quickly either establish a new outgoing connection. 
If $P_i$ were to close an outgoing connection,
then $P_k$ would only be able to wait for a new incoming connection request.
An additional useful heuristic (which we do not currently employ in our
simulations) when selecting connections to close would be to never close a
connection to a peer that is currently unchoked or is actively sending data.

Our preemption strategy assumes that $P_i$ somehow knows whether $P_j$ has
received its address from the tracker. The easiest way to implement
this functionality is to set a specific bit in the BitTorrent
HANDSHAKE message sent from $P_j$ to $P_i$. There are unused reserved
bits in the HANDSHAKE message that can be used for this purpose. As
$P_j$ is untrusted, $P_i$ will never accept more than a few percents
of preempted peers, typically $10\%$ of the peer set (yet in our 
evaluation, we put no limit on the number of accepted preempted peers). 
This way, a 
misbehaving or evil peer will not be able to harm a regular peer by
making him drop all its connections using preemption with fake peers.

Consequently, to implement the preemption strategy, one only needs to
modify clients, but not the tracker. Moreover, as this new strategy is
based on a specific bit set in the HANDSHAKE message, it is backward
compatible with existing BitTorrent clients. Indeed, the default behavior
of a BitTorrent client that receives a HANDSHAKE with an unknown bit
set is to ignore this bit. 

\section{Methodology}
\label{sec:methodology}

Before presenting our results, we first outline our experimental setup
and describe the simulation parameters. 
We then characterize the peer arrival and departure distributions we consider in this study, 
and present the metrics used to evaluate the properties of the overlay.

\subsection{Experimental Setup}

In order to investigate the properties of the overlays generated by the two strategies,
we developed a simulator that captures the evolution of the overlay structure over time 
as peers join and leave the torrent. The simulator source code is publicly
available~\cite{tracker_simulator_link}, and it follows the protocol as it is implemented 
in the official BitTorrent client version 4.0.2. 
We do not model the peer and piece selection strategies used in data exchange,
since we focus on the construction and robustness properties of the overlay instead.

We believe that simulations, rather than physical experiments, are a more appropriate 
vehicle for evaluating these properties, for three main reasons.
First, the BitTorrent overlay cannot be explored using a crawler, as is the case
for other peer-to-peer systems, such as Gnutella~\cite{STUT05_SIGMETRICS}. 
This is because the protocol itself does not offer a generic distributed mechanism 
for peer discovery, i.e., there is no way to make a BitTorrent peer (that does not support
the peer exchange extension) to provide any information about the peers in its peer set.
As several BitTorrent clients do not support this extension, the information we 
would get from public torrents would be largely incomplete. 

Second, we cannot analyze existing traces collected at various trackers,
since a peer never shares with the tracker its connectivity 
with other peers in the swarm.

Lastly, we could instead set up our own controlled testbed, e.g., on PlanetLab, running 
real experiments and collecting statistics. However, running such
experiments is harder and more time consuming than running
simulations, and it will not bring significantly
more insights than simulations. Indeed, a frequent argument against BitTorrent simulations, namely the
fact that it is challenging to accurately model the system dynamics,
is arguably not applicable in our case, since we focus exclusively on
the overlay construction, which is far easier to model than BitTorrent's data exchange. 

In any case, we have validated our simulation results by comparing
them against results from real experiments on a controlled
testbed. These experiments are not presented here due to space
limitations, but can be found in our technical
report~\cite{HAMR07_TECH}.

\subsection{Arrival and Departure Distributions}
\label{sec:simulation-parameters}

Following observations by Guo \textit{et al.}~\cite{GUO05_IMC}, we model peer
arrivals and departures with an exponential distribution. We split
simulated time into \textit{slots}. Slot $i$, where $i>=1$, is defined as the simulated
time elapsed between time $10 \cdot (i-1)$ minutes and 
time $10 \cdot i$ minutes. We focus on a flash crowd scenario, where
most peers arrive soon after the beginning of the torrent's lifetime. Thus,
within each time slot $i$, the number of new peers that join the
torrent is $ 1000\cdot \exp^{-0.7 \cdot (i-1)} ~if~i \leq 4$ and 
$0 ~if~i > 4$.  Each peer stays connected to the torrent for a random
period of time uniformly distributed between $10$ and $20$ simulated
minutes. Under these assumptions, $1000$ peers will arrive during the
first $10$ minutes, $497$ peers during the next $10$ minutes, $247$
peers between the $20^{th}$ and $30^{th}$ minute, and the remaining $123$ peers 
during the fourth 10-minute period. No peers will arrive after the first $40$
minutes of the simulation.  As a result, there will be more peer arrivals
than departures during the first two time slots, and vice versa
starting from the third time slot. The evolution of the torrent size
that results from this model corresponds to a typical real torrent,
based on previous studies~\cite{GUO05_IMC,izal04}.

The typical lifetime of a BitTorrent peer is in the order of several
hours, while the torrent lifetime ranges from several hours to several
months. In our simulations, the average peer and torrent lifetimes are around
$15$ and $70$ simulated minutes respectively. As we only focus on the
overlay construction during the flash crowd, considering longer
lifetimes would not give any new insights. Indeed, as our results show,
the peer arrival and departure order have a significant impact on the overlay, 
unlike the duration of their presence in the torrent.

\subsection{Evaluation Metrics}
We use three simple metrics to evaluate the structure of an overlay,
which we believe capture different important overlay properties well.
First, the \textit{bottleneck index} is defined as the ratio of the
number of connections between the first $80$ peers (equal to the maximum peer set size) 
to join the torrent (including the initial seed) and the rest of the peers, over the maximum
possible number of such connections ($80*80=6400$). This index provides an
indication of the presence of a bottleneck between the first set of participating peers
and the rest of the torrent. The existence of such a bottleneck would arguably adversely 
impact both the content distribution speed and robustness of the overlay. 
Note that a lower bottleneck index implies a worse bottleneck.

The second metric we use is the \textit{average peer set size}. A larger average peer 
set size implies a larger number of neighbors, which should lead to more opportunities 
of finding a peer that is willing to exchange data and higher resilience to churn.

Lastly, we measure the \textit{overlay diameter} as the maximum number of hops in the torrent. 
A small diameter indicates that a piece can reach any peer within a few hops.
Therefore, this metric also serves to evaluate the diversity of pieces in the system, which 
has been shown to lead to efficient piece replication~\cite{lego06_IMC}.

\begin{figure*}[t]
\centering
\subfigure[Bottleneck Index]{\includegraphics[width=0.3\textwidth]
{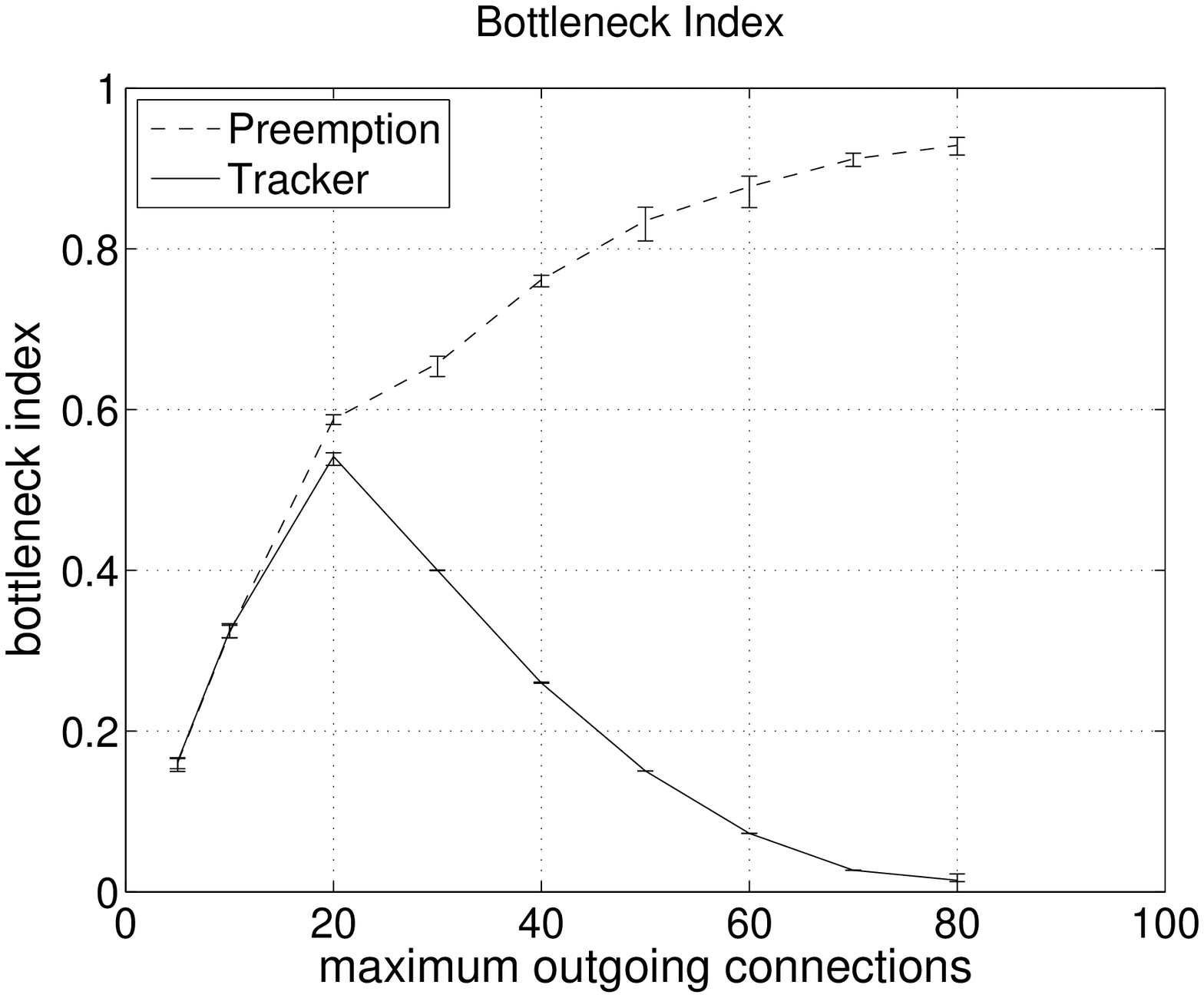}
\label{fig:all-metrics-bottleneck}}
\hfill{}
\subfigure[Average Peer Set Size]{\includegraphics[width=0.3\textwidth]
{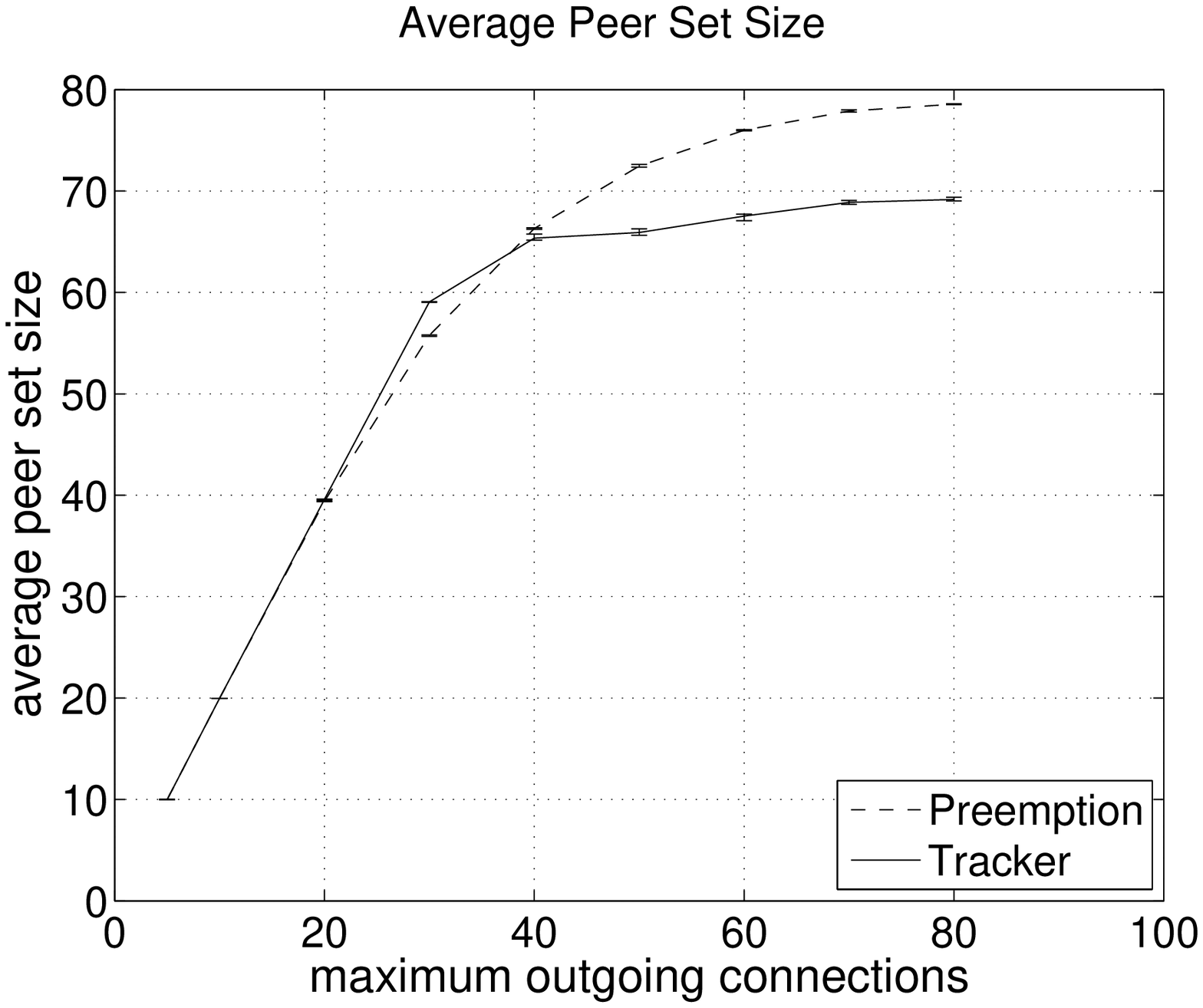}
\label{fig:all-metrics-average}}
\hfill{}
\subfigure[Overlay Diameter]{\includegraphics[width=0.3\textwidth]
{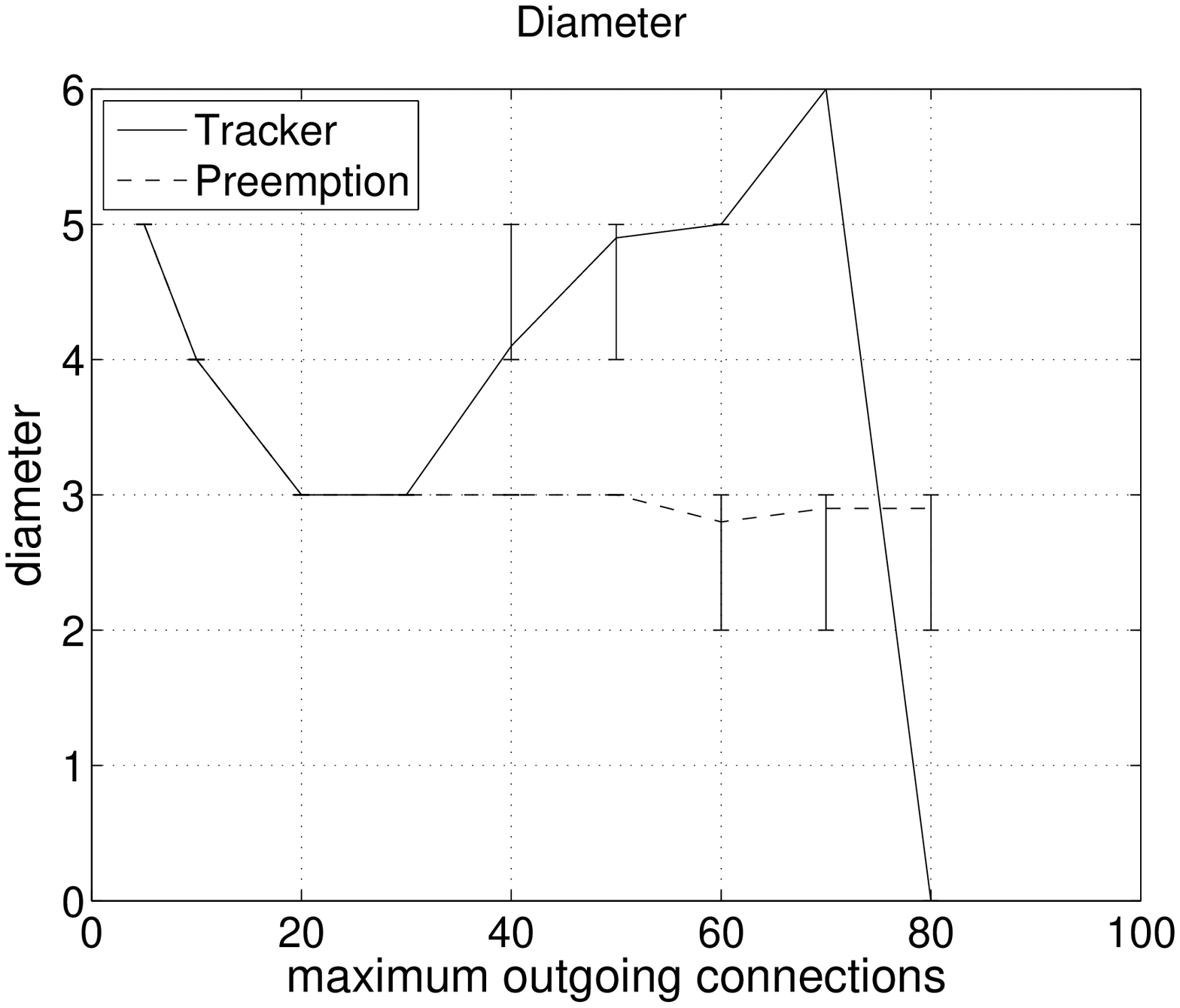}
\label{fig:all-metrics-diameter}}
\hfill{}
\caption{Bottleneck index, average peer set size, and overlay diameter as a
  function of the \maxoc{}, averaged over ten independent runs.  The
  error bars indicate the minimum and maximum over all runs.
  {\textit{There is no single value of $O_{max}$ that optimizes all three
  metrics for the default tracker strategy. Moreover, the preemption strategy 
  outperforms the default one for all three metrics.}}}
\label{fig:all-metrics}
\end{figure*}

\begin{figure*}
\centering
\subfigure[$O_{max}=20$]{\includegraphics[width=0.23\textwidth]
{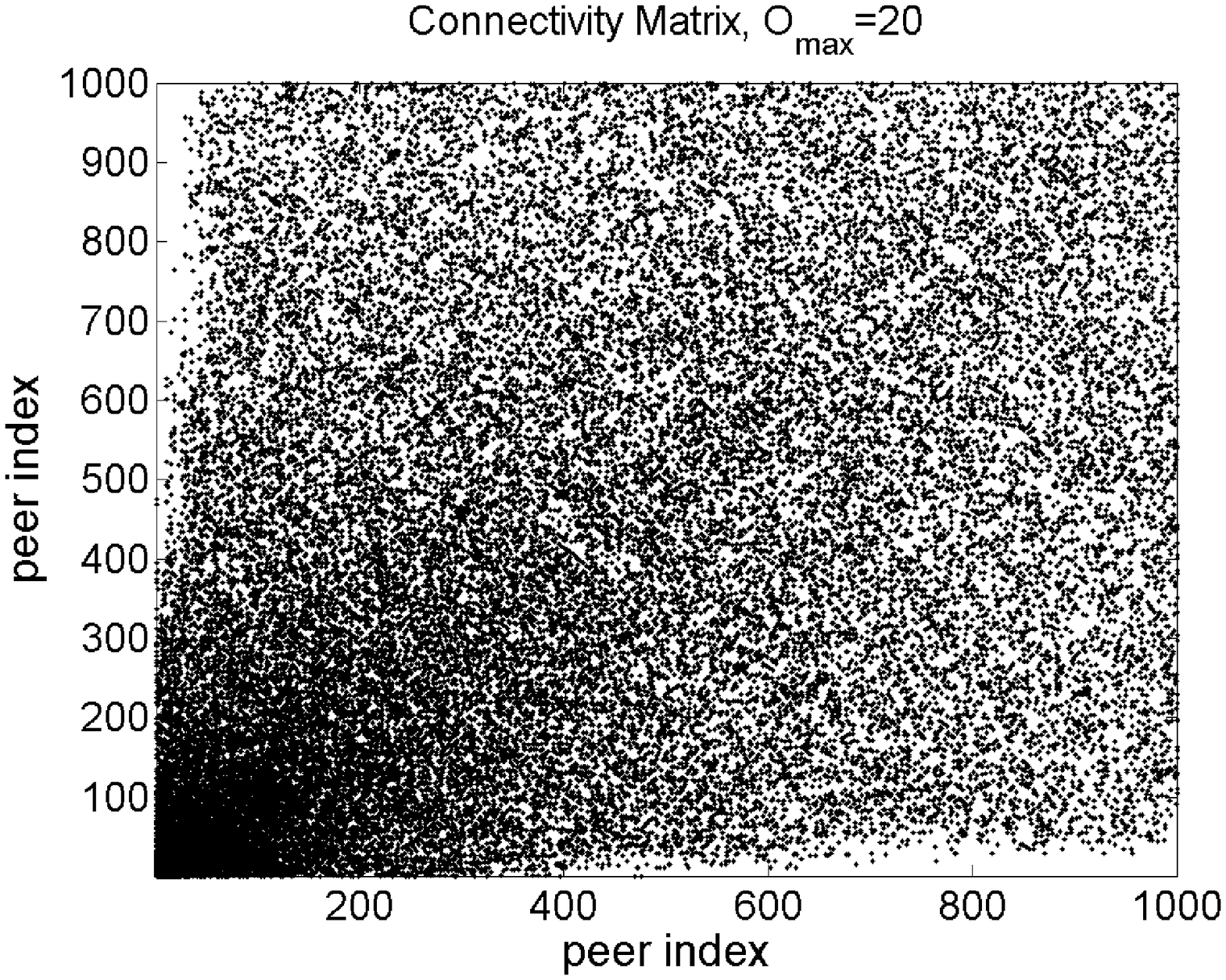}}
\hfill{}
\subfigure[$O_{max}=40$]{\includegraphics[width=0.23\textwidth]
{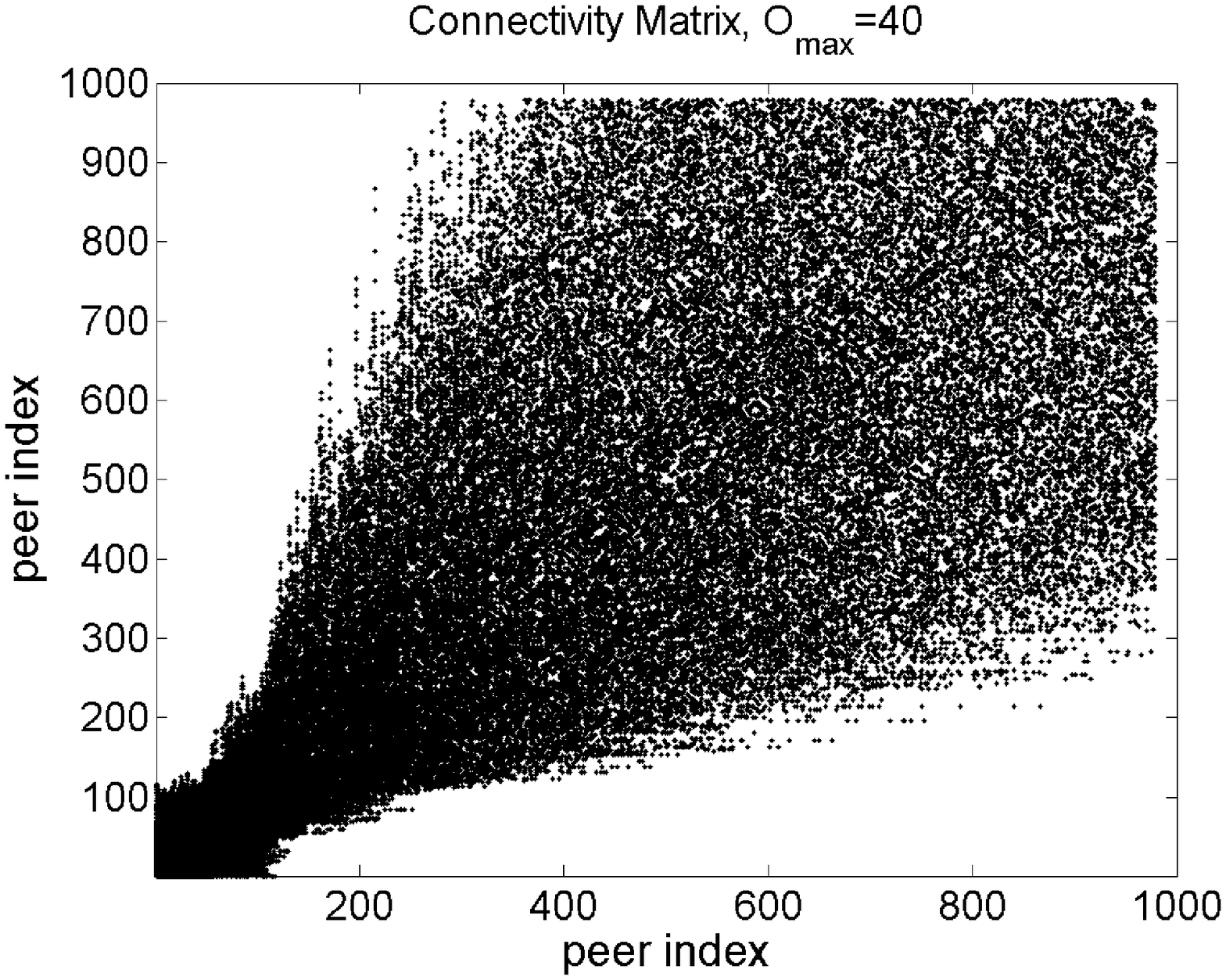}
\label{fig:connectivity-nopreempt-40}}
\hfill{}
\subfigure[$O_{max}=60$]{\includegraphics[width=0.23\textwidth]
{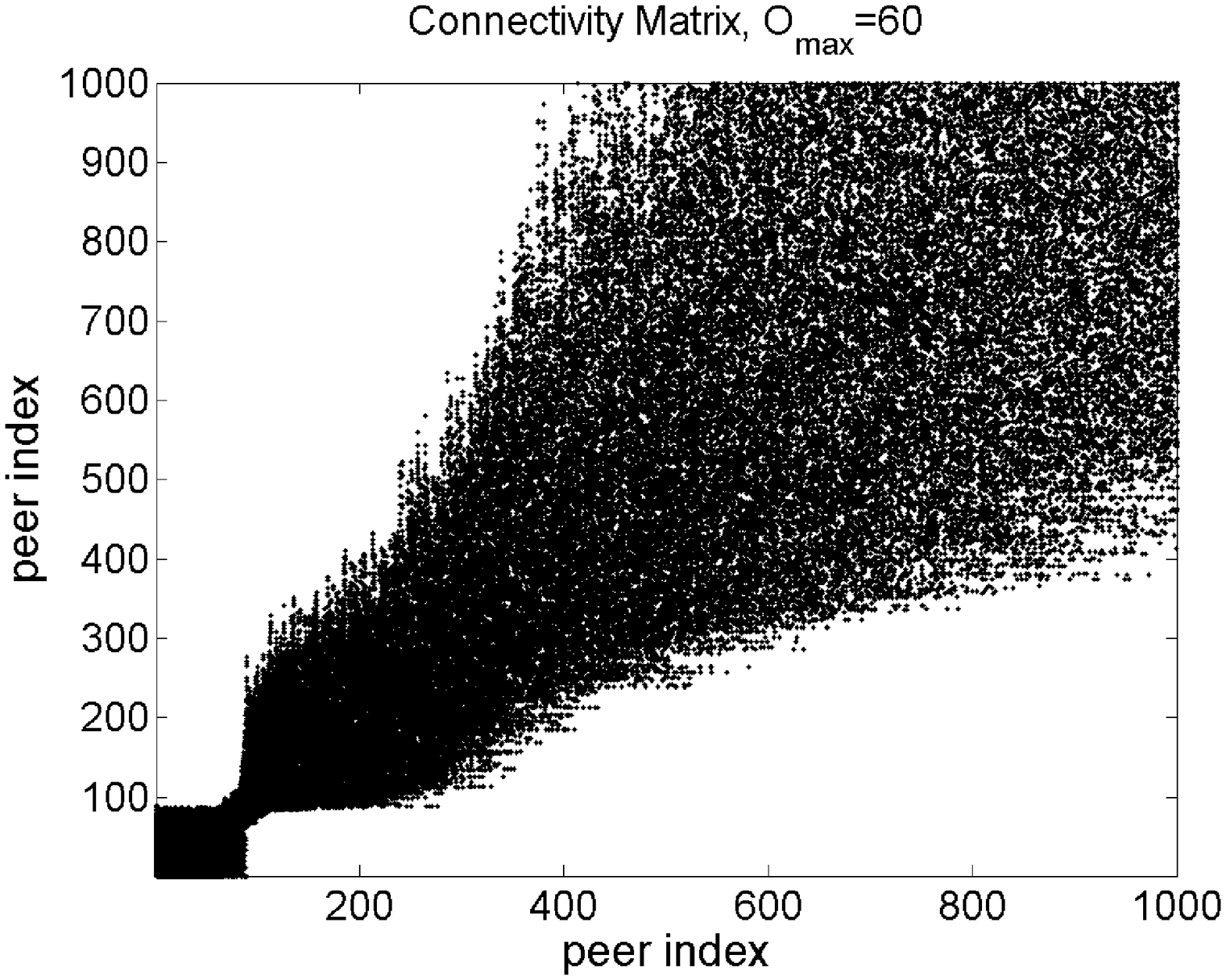}
\label{fig:connectivity-nopreempt-60}}
\hfill{}
\subfigure[$O_{max}=80$]{\includegraphics[width=0.23\textwidth]
{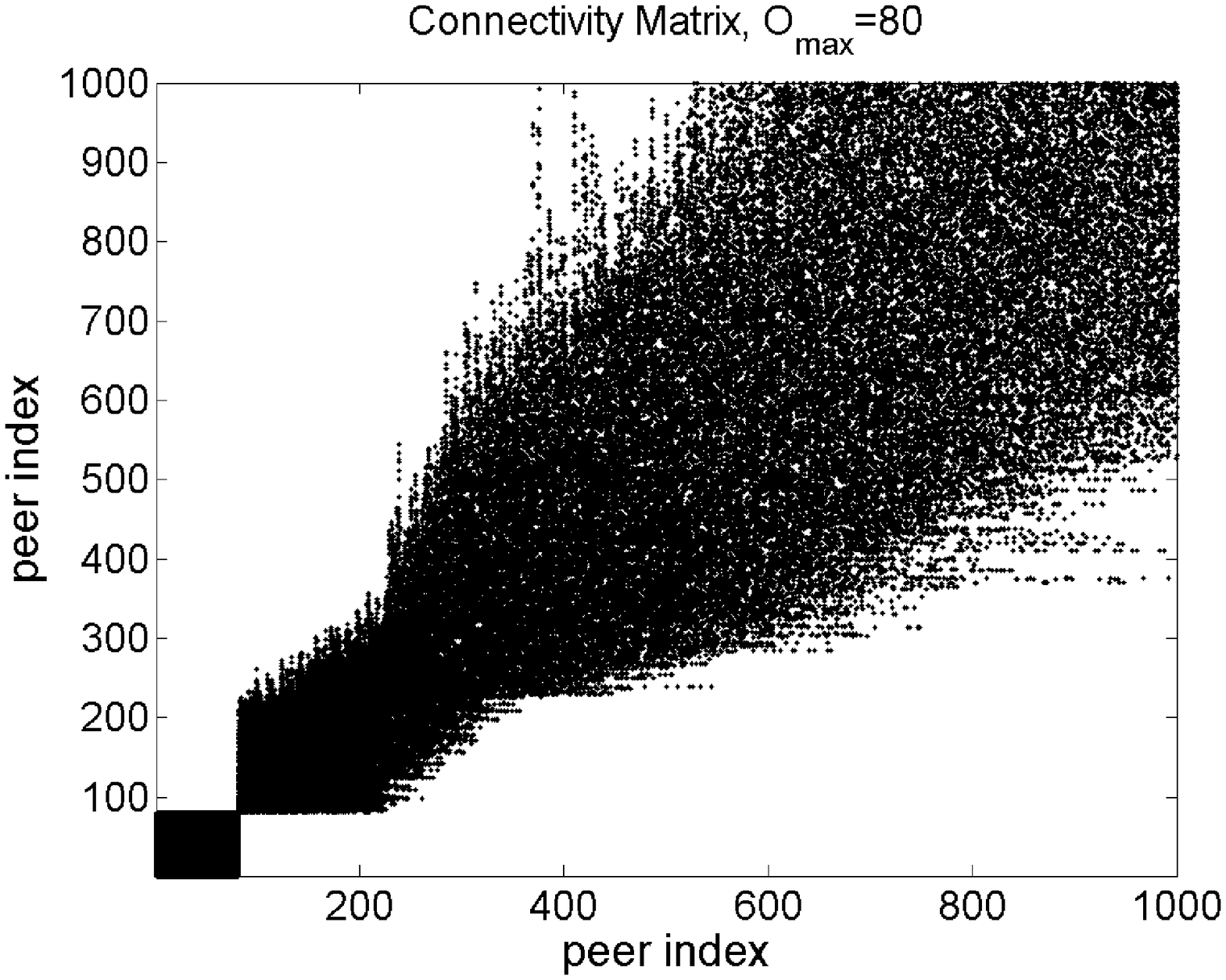}}
\caption{Connectivity matrices for the overlay generated by the tracker
  strategy after $10$ minutes, for a single representative run. 
  A dot at (i,j) means that i and j are neighbors.
  \emph{The tracker strategy generates a graph that suffers from a bottleneck 
   depending on the value of $O_{max}$; the larger $O_{max}$ is, the more clustered 
   the first $80$ peers become.}}
\label{fig:connectivity-nopreempt}
\end{figure*}

\section{Simulation Results} 
\label{sec:results}

\begin{figure*}[t]
\centering
 \subfigure[$O_{max}=20$]{\includegraphics[width=0.23\textwidth]
 {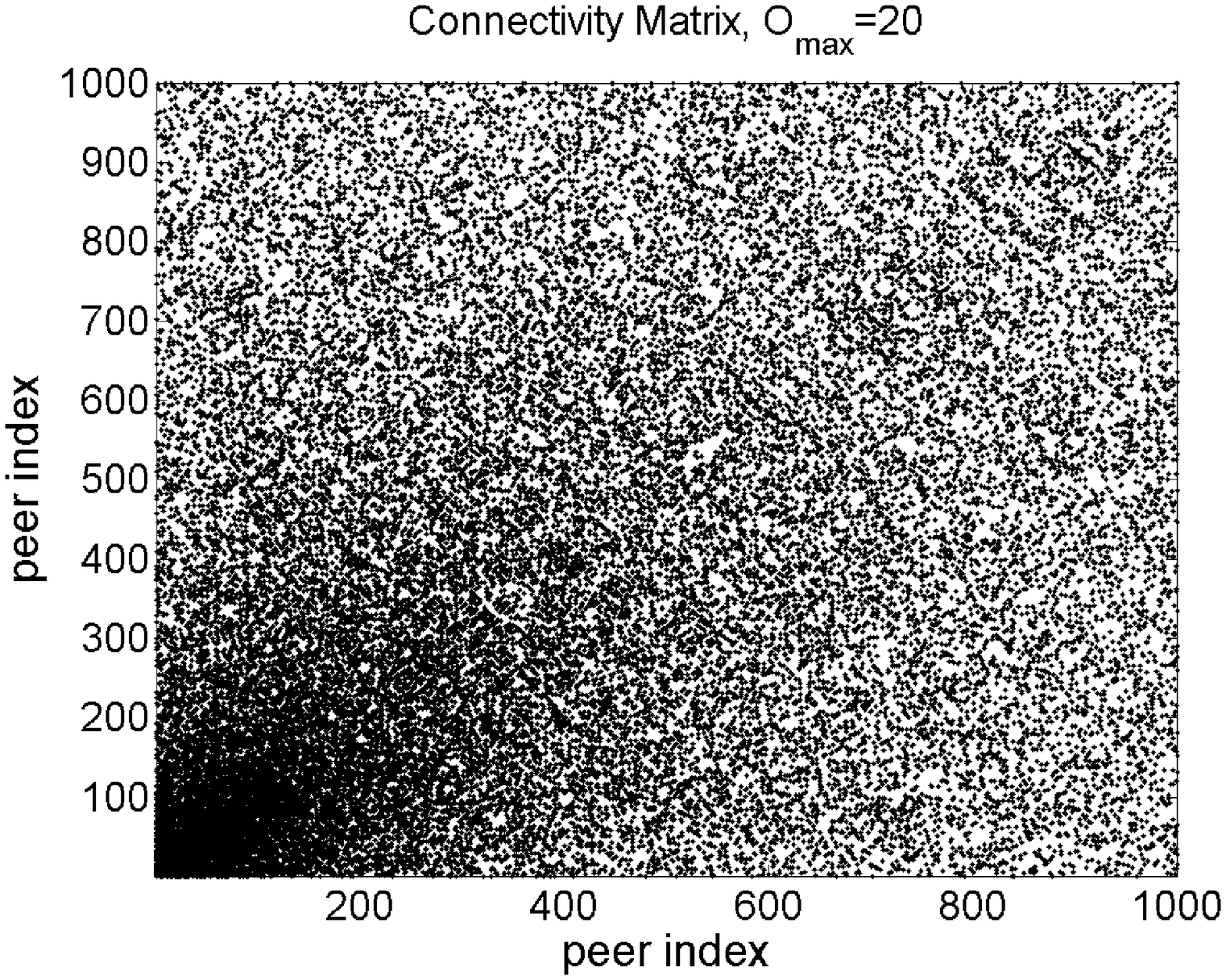}}
 \hfill{}
\subfigure[$O_{max}=40$]{\includegraphics[width=0.23\textwidth]
{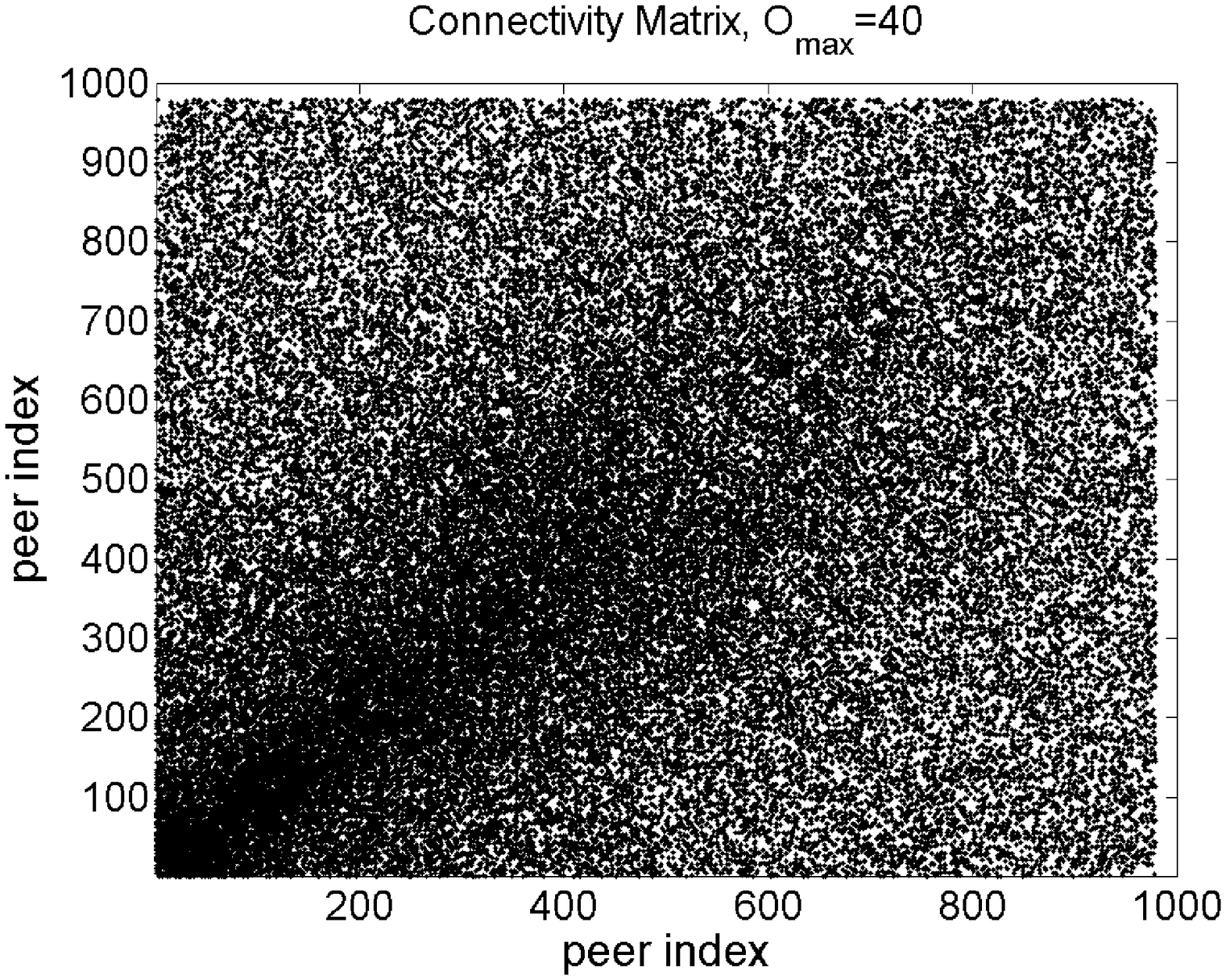}}
\hfill{}
 \subfigure[$O_{max}=60$]{\includegraphics[width=0.23\textwidth]
 {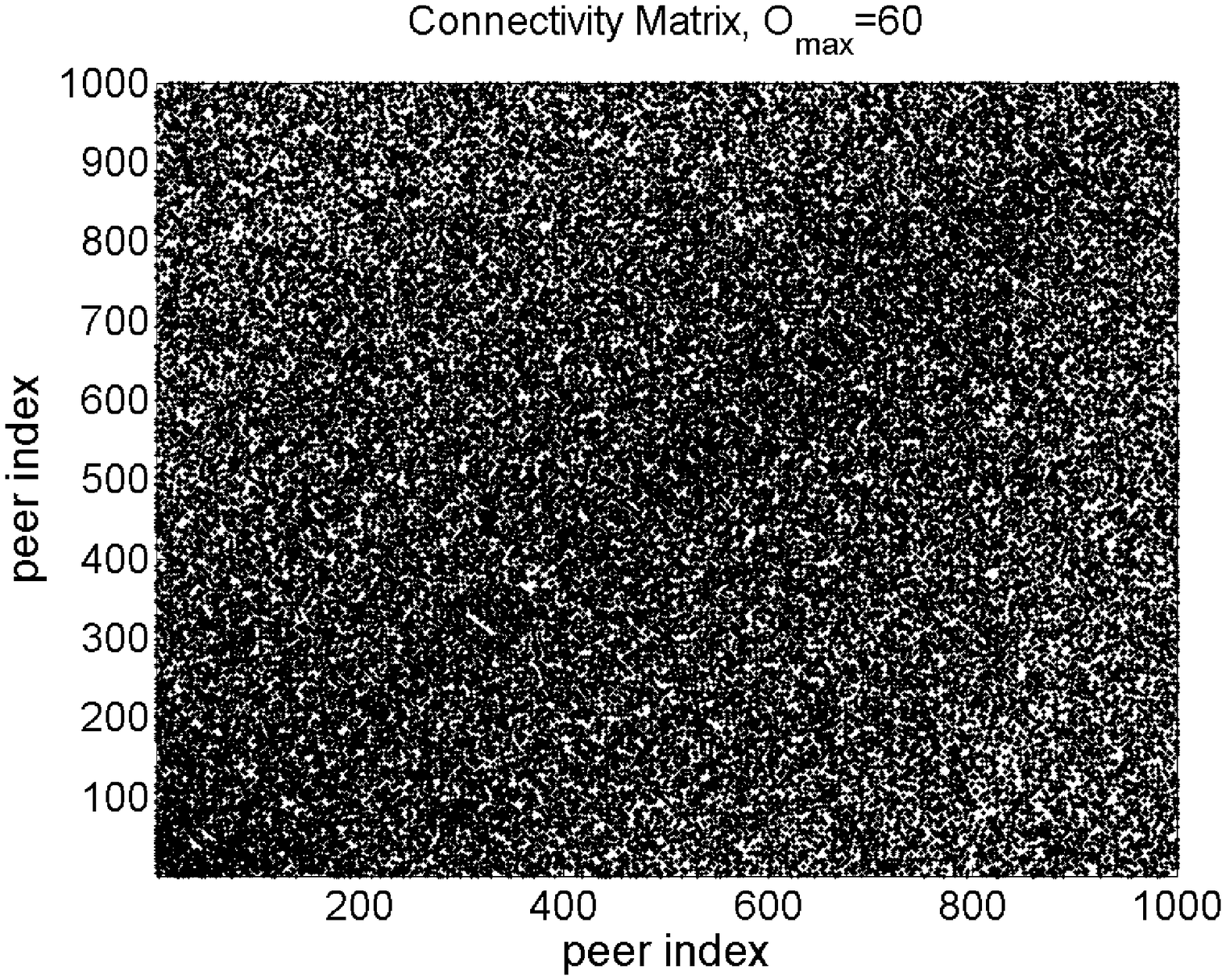}}
 \hfill{}
\subfigure[$O_{max}=80$]{\includegraphics[width=0.23\textwidth]
{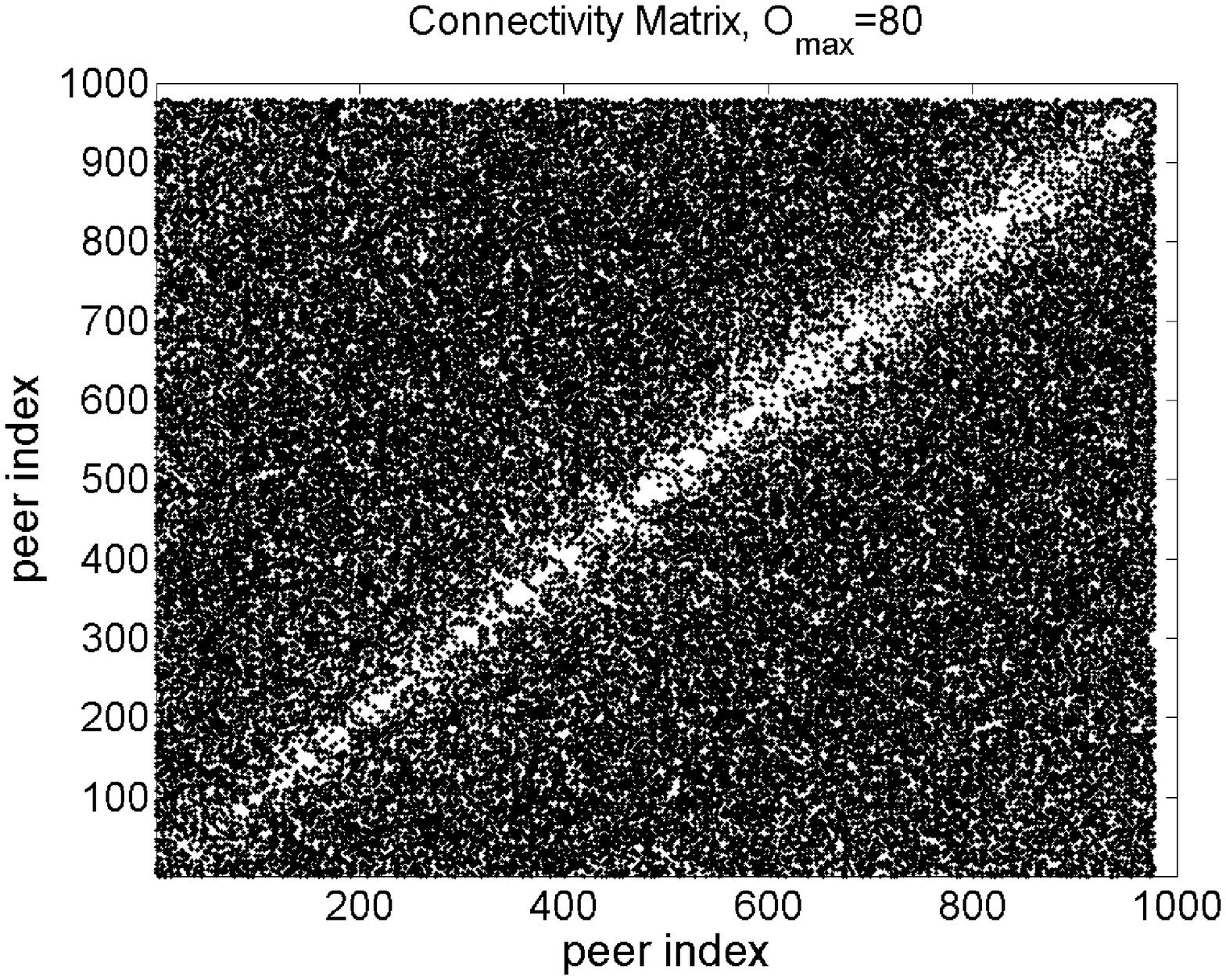}}
\caption{Connectivity matrices for the overlay generated by the
  preemption strategy after $10$ minutes, for a single representative run. 
  A dot at (i,j) means that i and j are neighbors. 
  \emph{We no longer observe any clear clustering among the peers who joined first.}}
\label{fig:connectivity-preempt}
\end{figure*}

In our simulations, we use the official BitTorrent client's default parameter values
and set the maximum peer set size to
$80$ and the minimum number of neighbors to $20$. We then vary the maximum
number of outgoing connections $O_{max}$ from $5$ to $80$ with a step
of $5$. 

We evaluate the properties of overlays generated using both the default and the
preemption strategies.
Figure~\ref{fig:all-metrics} plots the three metrics we consider over the 
maximum number of outgoing connections.
We observe that, for the tracker strategy (solid line), there is no value 
of $O_{max}$ that optimizes all three metrics. The highest bottleneck index, which
would result in a more robust overlay,
as well as a relatively small overlay diameter are both achieved for $O_{max}$ around
$20$. However, the optimal average peer set size occurs for $O_{max}$ equal to $80$.
As a result, the common practice to set $O_{max}$ to half of the
maximum peer set size (thus making it $40$) is by no means the best choice. Rather, 
since average peer set size approaches its maximum for $O_{max}$ equal to $30$, 
we propose setting the maximum number of outgoing connections between $20$ and $30$, 
which achieves a better trade-off for the three metrics we consider.

In addition, we see in Figure~\ref{fig:all-metrics-diameter} that the overlay diameter 
for the tracker strategy is $0$ when $O_{max}$ is set to $80$. This means that the peer 
graph is partitioned into two separate subgraphs. We also observe that the bottleneck 
index increases for $O_{max}$ values up to $20$ and decreases for larger values. 
To explain these results we focus now on the actual connections among peers in the overlay. 
Figure~\ref{fig:connectivity-nopreempt} plots these connections for the tracker strategy, 
captured after $10$ simulated minutes, i.e., after the arrival of $1000$ peers (see
Section~\ref{sec:simulation-parameters}), for four distinct values of
$O_{max}$: $20$, $40$, $60$, and $80$. 
The results are shown in the form of a \textit{connectivity matrix}, where a dot at
$(i, j)$ means that peers $i$ and $j$ are neighbors.

We observe that, for lower values of $O_{max}$, there exists good connectivity among 
peers, with some clustering being observed for those who joined the torrent first. 
However, when increasing $O_{max}$ further, we see the formation of a small 
cluster that consists of the first $80$ peers (same as the maximum peer set size). 
This clustering becomes clearer for increasingly larger values of $O_{max}$, to the point 
where, for $O_{max}$ equal to the maximum peer set size, those first $80$ peers form 
a completely separate partition from the rest of the overlay. The creation of two separate
partitions will definitely be harmful to system robustness, as the seed and the first 
peers, who already have most of the pieces, will be unable to share them with the rest of 
the system.

We attempt to explain the reason behind this clustering with an
example. In the following, peer $P_{i}$ is the $i^{th}$ peer to join
the torrent. 
For $O_{max}$ equal to $40$ (see Fig.~\ref{fig:connectivity-nopreempt-40}), all of peer 
$P_{25}$'s neighbors belong to the first $100$ peers who joined the torrent. The reason is
that when peer $P_{25}$ arrives, it establishes outgoing connections to all other $24$ peers 
already in the system. It then waits for new arrivals in order to establish the remaining 
$56$ connections it still needs to reach its maximum peer set limit. Those missing connections 
are satisfied after the arrival of another $75$ peers on average, $P_{26}, \dots, P_{100}$. 
Similarly, when peer $P_{200}$ arrives, it establishes up to $40$ outgoing connections.  
However, $P_{200}$ now needs to wait for the arrival of a larger number of peers in order 
to establish its remaining $40$ incoming connections, because the probability that its IP 
address is returned by the tracker to new peers decreases as the number of peers in the 
torrent increases. This explains why, as compared to $P_{25}$, the neighbors of $P_{200}$
belong to a larger set of peers ($P_{60}, \dots, P_{600}$).

This leads us to believe that an alternative strategy that introduces some randomness
into the connection establishment process would exhibit better behavior.
With that in mind, let us now look at the properties of an overlay built using our proposed
preemption strategy. As shown in Figure~\ref{fig:all-metrics} (dashed line), such an overlay
exhibits better characteristics than the one generated with the tracker strategy, for the three
metrics we consider.
Moreover, a value $O_{max}$ equal to the maximum peer set size clearly gives the best results 
for all three metrics. In addition, looking at the connectivity matrices of the overlay built 
with the preemption strategy captured after $10$ simulated minutes (shown in 
Figure~\ref{fig:connectivity-preempt}), we observe good connectivity among peers for all 
value of $O_{max}$, without the clustering effects observed with the tracker strategy.

Thus, the preemption strategy obviates the need to heuristically select the maximum
number of outgoing connections allowed at each peer, as the best overlay structure 
is always attained when that parameter is equal to the maximum peer set size.
In addition, the proposed strategy outperforms the default one for all considered metrics.
Therefore, the preemption strategy, while simple and easy to implement, offers a strong
alternative to the one used by most BitTorrent clients.

\section{Related Work}
\label{sec:related}

There has been a fair amount of work on the performance and robustness of 
BitTorrent systems, most of which is complementary to ours.

Bram Cohen, the protocol's creator, first described its main mechanisms and their 
design rationale~\cite{cohen03}. 
Several measurement studies attempted to characterize the protocol's properties by
examining real BitTorrent traffic.
Izal \textit{et al.}~\cite{izal04} measured several peer characteristics derived
from the tracker log of the Red Hat Linux 9 ISO image, including the proportion of 
seeds and leechers and the number and geographical spread of active peers.
They observed that, while there is a correlation between upload and download rates, 
the majority of content is contributed by only a few leechers and the seeds.

Pouwelse \textit{et al.}~\cite{pouwelse05} studied the content availability, 
integrity, and download performance of torrents on a once-popular tracker website.
Andrade \textit{et al.}~\cite{andrade05} additionally examined BitTorrent 
sharing communities and found that sharing-ratio enforcement and the use of RSS 
feeds to advertise new content may improve peer contributions. 
At the same time, Guo \textit{et al.}~\cite{GUO05_IMC} demonstrated that the rate of 
peer arrival and departure from typical torrents follows an exponential distribution 
and that performance fluctuates widely in small torrents.
They also proposed inter-torrent collaboration as an incentive for leechers to stay 
connected as seeds after the completion of their download.
A more recent study by Legout \textit{et al.}~\cite{lego06_IMC} examined peer behavior 
by running extensive experiments on real torrents. 
They showed that the rarest-first and choking algorithms play a critical role in 
BitTorrent's performance, and claimed that the use of a volume-based tit-for-tat 
algorithm, as proposed by other researchers~\cite{jun05}, is not appropriate. 

There have also been some simulation studies attempting to better understand BitTorrent's 
system properties.
Felber \textit{et al.}~\cite{felber04} performed an initial investigation of the 
impact of different peer arrival rates, peer capacities, and peer and piece selection 
strategies.
Bharambe \textit{et al.}~\cite{bharambe06} utilized a discrete event simulator to 
evaluate the impact of BitTorrent's core mechanisms and observed
that rate-based tit-for-tat incentives cannot guarantee fairness.
They also showed that the rarest-first algorithm outperforms alternative piece 
selection strategies. 
Lastly, Tian \textit{et al.}~\cite{TIAN06_INFOCOM} studied peer performance toward the 
end of the download and proposed a new peer selection strategy that enables more 
peers to complete their download, even after the departure of all the seeds.

Our work differs from all previous studies in its approach and results. 
We performed extensive simulations to examine the impact of the overlay construction
strategy on system properties and robustness.
Our results showcase the importance of the maximum number of outgoing connections
and propose a concrete improvement to the protocol.

\section{Summary and Future Work}
\label{sec:conc}

In this paper, we introduce a new preemptive overlay construction strategy for BitTorrent.
We evaluate it along with the default BitTorrent tracker strategy for a flash crowd 
scenario, based on three different metrics. Our results show that the tracker strategy 
is quite sensitive to the maximum number of outgoing connections, which does not seem 
to have a single value that optimizes all metrics. In addition, a value between $20$ and 
$30$ offers a better choice than the current BitTorrent default of $40$. On the other
hand, the proposed preemption strategy outperforms the default one for all three 
metrics considered. Furthermore, there is a clear optimal choice for the maximum number 
of outgoing connections (equal to the maximum peer set size), a fact that removes the need
to set this parameter in an ad-hoc manner.

These results already provide some initial insights into how the default tracker
strategy behaves and how to improve it using preemption. However, many questions remain 
open for future work. 
First, while we have introduced specific metrics for evaluating the properties of the
overlay structure, and we have discussed how these metrics are linked to the system's 
robustness, we have not formally quantified their impact. This is a necessary step in
understanding how the overlay structure actually affects system
properties. In addition, it would be interesting to investigate whether the
preemption strategy can be exploited by an attacker in order to
disconnect peers from torrents. 

Finally, while we have examined the tracker strategy and its preemption-based 
alternative, there exist other strategies based on gossiping, e.g., peer exchange, which 
are also promising. Some preliminary results in that 
direction show that such strategies produce an overlay with large diameter and low
bottleneck index, but they achieve the best average peer set size.
It would be interesting to better understand the trade-offs involved in such gossiping techniques,
and incorporate some of their features into our preemption strategy.

\section*{Acknowledgments}
We want to thank 
Matthieu Latapy for his suggestions on the use of preemption, and Christos 
Gkantsidis for his helpful comments.


\end{document}